\documentclass[aps,prd,preprint,superscriptaddress,showpacs,floatfix,
preprintnumbers,nofootinbib,a4paper]{revtex4-1}
\usepackage{hyperref}
\usepackage{color}
\usepackage{graphicx}
\usepackage{textcomp}
\usepackage{subfigure}
\usepackage{amsmath,amssymb}
\usepackage{enumerate}
\newcommand{\mathsym}[1]{{}}

\newcommand{\ba}{\begin{array}}
\newcommand{\ea}{\end{array}}
\newcommand{\be}{\begin{equation}}
\newcommand{\ee}{\end{equation}}
\newcommand{\beqa}{\begin{eqnarray}}
\newcommand{\eeqa}{\end{eqnarray}}
\def\mt{$\mu$-$\tau$~}
\newcommand{\group}[1]{\textlbrackdbl #1\textrbrackdbl}

\begin{document}
\vspace*{1cm}
\title{Discrete flavour symmetries for degenerate solar neutrino pair and their predictions}
\bigskip
\author{Anjan S. Joshipura}
\email{anjan@prl.res.in}
\affiliation{Physical Research Laboratory, Navarangpura, Ahmedabad 380 009, India.}
\author{Ketan M. Patel}
\email{ketan.patel@pd.infn.it}
\affiliation{Istituto Nazionale Fisica Nucleare, Sezione di Padova, I-35131 Padova,
Italy. \vspace*{1cm}}

\begin{abstract}
\vspace*{0.2cm}
Flavour symmetries appropriate for describing a neutrino spectrum with degenerate solar pair  
and a third massive or massless neutrino are discussed. We demand that the required residual
symmetries of the leptonic mass matrices be subgroups of some discrete symmetry group $G_f$. $G_f$
can be a subgroup of $SU(3)$ if the third neutrino is massive and we derive general results on the 
mixing angle predictions for various discrete subgroups of $SU(3)$. The main results are: (a) All
the $SU(3)$ subgroups of type C fail in simultaneously giving correct $\theta_{13}$ and
$\theta_{23}$. (b) All the groups of type D can predict a relation $\cos^2\theta_{13}
\sin^2\theta_{23}=\frac{1}{3}$ among the mixing angles which appears to be a good zeroth order
approximation. Among these, various $\Delta(6n^2)$ groups with $n\geq 8$ can simultaneously lead
also to $\sin^2 \theta_{13}$ in agreement with global fit at 3$\sigma$. (c) The group
$\Sigma(168)\cong PSL(2,7)$ predicts near to the best fit value for $\theta_{13}$ and $\theta_{23}$
within the 1$\sigma$ range. All discrete subgroups of $U(3)$ with order $<512$ and having three
dimensional irreducible representation are considered as possible $G_f$ when the third neutrino is
massless. Only seven of them are shown to be viable and three of these can correctly predict
$\theta_{13}$ and/or $\theta_{23}$. The solar angle remains undetermined at the leading order in all
the cases  due to degeneracy in the masses. A class of general perturbations which can correctly
reproduce all the observables are discussed in the context of several groups which offer
good leading order predictions. 
\end{abstract}

\maketitle

\section{Introduction}
The lack of direct evidence of the neutrino mass scale allows three distinct possibilities for the
neutrino masses, hierarchical, quasidegenerate or partially degenerate spectrum consisting of two
degenerate states forming a solar pair in zeroth order and the third massive or nearly massless
state. Mass of the third state (solar pair) lies at the atmospheric scale in case of the normal
(inverted) hierarchy in neutrino masses. The third neutrino can even be massless in case of the
inverted hierarchy. Near degeneracy of two masses has an economical explanation in which two of the
active neutrinos combine to form a Dirac state with very tiny mass difference between them. This
requires special symmetries of the neutrino mass matrix and eventually of the underlying Lagrangian
if this symmetry is not accidental. There exists numerous examples of continuous symmetries starting
from minimal $U(1)$, e.g. $L_e-L_\mu-L_\tau$ (see \cite{Mohapatra:2005wg} for an exhaustive
reference list) to larger groups $O(3)_l\times O(3)_e\times O(3)_\nu\times U(1)_R$
\cite{Joshipura:2009gi} which lead to two or all three degenerate neutrinos. But discrete symmetries
achieving this are not much explored except in some simple cases, e.g. $S_3$ \cite{Jora:2006dh},
$A_4$ \cite{Ma:2001dn}. Aim of the present paper is to make an exhaustive search for discrete
symmetries leading to degenerate solar pair  and work out their predictions  not only for the
neutrino mass pattern but also for the leptonic mixing angles and CP phases.

Method of searching for possible symmetry groups $G_f$ of an underlying theory from the symmetries
of the leptonic mass matrices is discussed and studied extensively in the literature, see
\cite{Altarelli:2010gt,Altarelli:2012ss,Smirnov:2011jv,King:2013eh,Ishimori:2010au} for recent
reviews. It is assumed that symmetries $G_\nu$ and $G_l$ of the neutrino and the charged lepton mass
matrices respectively are not accidental but arise when $G_f$ is spontaneously broken. In this case,
if $G_\nu$ and $G_l$ are known a priori then the minimal group $G_f$ which contain these as
subgroups may be regarded as a symmetry of the theory. This approach makes definite prediction for
the leptonic mixing matrix which solely depend on the choice of $G_\nu$ and $G_l$
\cite{Lam:2007qc,Lam:2008rs,Lam:2008sh,Lam:2012ga,Lam:2011ag} while dynamics is invoked to assure
that $G_f$ is broken down to the required symmetries of the leptonic mass matrices. Possible choices
of $G_f$ based on this approach and the resulting mixing patterns are studied extensively in case of
the three non-degenerate massive Majorana neutrinos
\cite{Toorop:2011jn,deAdelhartToorop:2011re,Hernandez:2012ra,Hernandez:2012sk, Holthausen:2012wt,
Parattu:2010cy,Fonseca:2014koa} and also Dirac neutrinos \cite{Holthausen:2013vba,Hagedorn:2013nra}.
Using a similar approach, a new class of $G_f$ is proposed and extensively studied recently by us in
case of Majorana neutrinos with a massless state \cite{Joshipura:2013pga,Joshipura:2014pqa}. Our aim
is to apply the same method for the description of the neutrino mass spectrum with an active Dirac
pair and a massive or massless Majorana neutrino. The required residual symmetries and the nature of
the resulting Dirac neutrino is quite different from the case of the conventional Dirac neutrinos
studied earlier in \cite{Holthausen:2013vba,Hagedorn:2013nra}. The masses of the conventional  Dirac
neutrinos are not naturally suppressed due to inherent lepton number conservation 
and one needs to invoke some additional mechanism to suppress them. In contrast,
at least some of the lepton numbers $L_e,L_\mu,L_\tau$  are broken if not all when two of the active
neutrinos combine to form a Dirac state and the smallness of neutrino masses gets directly linked to
lepton number violation.  The earlier study of the active Dirac pair which also included
the case of all degenerate neutrinos was presented in \cite{Hernandez:2013vya} but it was  
restricted only to the finite von-Dyck groups which accommodates the cyclic groups, dihedral groups,
$A_4$, $S_4$ and $A_5$ \cite{Hernandez:2012ra,Hernandez:2012sk}. The following analysis goes
well beyond these groups and encompasses all the  discrete subgroups (DSG) of $SU(3)$ having three
dimensional irreducible representations (IR) and leads to many new predictions for mixing angles. In
addition, we also explore the DSG of $U(3)$ which can be used as the symmetry of degenerate solar
pair and a massless neutrino.

In section \ref{residual}, we first briefly review the general approach of building $G_f$ from the
knowledge of leptonic mixing angles and then modify it to accommodated the possibility of degenerate
solar pair. Following it, we analyze different DSG of $SU(3)$ for their predictions of mixing angles
in section \ref{SU3-ana} and provide a numerical scan of such groups in section \ref{SU3-num}. We
then discuss the possibility of having a degenerate solar pair and massless third neutrino from DSG
of $U(3)$ in section \ref{U3} and provide some realistic examples of generating the solar mass
difference in section \ref{examples}. The study is finally summarized in section \ref{summary}.

\section{Residual symmetries leading to degenerate solar pair}
\label{residual}
Let us first briefly review the widely discussed approach
\cite{Lam:2007qc,Lam:2008rs,Lam:2008sh,Lam:2012ga,Lam:2011ag,Toorop:2011jn,deAdelhartToorop:2011re}
in which  the knowledge of lepton mixing
pattern is used to obtain the underlying leptonic symmetries. The Majorana neutrino
mass matrix $M_\nu$ with three arbitrary non-zero masses is known to be always invariant under
$Z_2\times Z_2$ group which can be defined in an arbitrary weak basis as 
\be \label{z2z2}
S_1=V_\nu~{\rm Diag.}(1,-1,-1)~ V_\nu^\dagger ~~{\rm and}~~ S_2=V_\nu~{\rm
Diag.}(-1,1,-1)~V_\nu^\dagger~,\ee
By construction, $S_1$ and $S_2$ commute. Likewise, the combination $M_lM_l^\dagger$ of the  charged
lepton mass matrix $M_l$ is invariant under a $Z_n\times Z_m\times Z_p$ symmetry
\be \label{talpha}
T_l=V_l~{\rm Diag.}(e^{i \phi_e},e^{i \phi_\mu},e^{i \phi_\tau})~V_l^\dagger, \ee
in an arbitrary basis with $\phi_{e,\mu,\tau}$ being some discrete phases. The predictive power of
these symmetries follow from the observation
\cite{Lam:2007qc,Lam:2008rs,Lam:2008sh,Lam:2012ga,Lam:2011ag,Toorop:2011jn,deAdelhartToorop:2011re}
that the neutrino mixing matrix $U_{\rm PMNS}$ is determined in terms of $V_\nu$ and $V_l$ : 
\be \label{upmns}
U_{\rm PMNS}= V_l^\dagger V_\nu~,\ee
$V_l$ and $V_\nu$ should be different
to get a non-zero mixing which means that the groups generated by $S_i$ and $T_l$ do not commute.
Thus they cannot be imposed as a symmetry in the basic Lagrangian. However, these groups can appear
as subgroups of some bigger symmetry $G_f$ whose breaking can lead to the leptonic mass matrices
invariant under respective symmetries. Extensive studies of possible $G_f$ and the resulting mixing
patterns \cite{Toorop:2011jn,deAdelhartToorop:2011re,Hernandez:2012ra,Hernandez:2012sk,
Holthausen:2012wt,Parattu:2010cy} show that the groups which can predict all three mixing angles
correctly within 3$\sigma$ are few and and big \cite{Holthausen:2012wt} but one can obtain several
good zeroth order mixing patterns from the smaller groups like, $A_5,~S_4$ etc.

The diagonal parts in definition of $S_{1,2}$ correspond to trivial symmetries of change of sign of
the neutrino fields in their mass basis. This leads to two possible generalizations which will be
considered in this paper. When one of the neutrinos is massless, then the phase of the corresponding
mass eigenstate can be changed without affecting the underlying Lagrangian \cite{Joshipura:2013pga}.
In this case, one of the $S_i$ is replaced by 
\be \label{snu12}
\tilde{S}_{\nu}= V_\nu ~{\rm Diag.}(\eta,1,-1)~ V_\nu^\dagger,\ee
with $\eta^n=1$ and $n\geq 3$. The $\tilde{S}_{\nu}$ forms a $Z_n$ ($Z_{2n}$) group for even (odd)
$n$. The $\tilde{S}_{\nu}$ thus represents a symmetry of a mass matrix with one massless and two
non-degenerate massive Majorana neutrinos. Since ${\rm det}(\tilde{S}_{\nu}) \neq \pm 1$, the group
embedding this would be a DSG of $U(3)$ rather than of $SU(3)$. This idea was proposed in
\cite{Joshipura:2013pga} and subsequently exhaustive search of possible DSG of
$U(3)$ was made and predictions for mixing angles were worked out in \cite{Joshipura:2014pqa}. It
was found that only one of the 75 DSG of $U(3)$ having order $<512$ and possessing 3-dimensional 
IR was able to predict a massless neutrino and three mixing angles
moderately close to their experimental values at the leading order.

A slight variation of the above two symmetries can be used to describe a pair of degenerate
neutrinos and a third massive or massless neutrinos. Consider the following discrete symmetry in
arbitrary basis 
\be \label{snu}
S_{1\nu}=V_\nu~{\rm ~Diag.}(\eta,\eta^*,1)~V_\nu^\dagger,\ee
where $\eta^n=1$ and $n\geq3$. The neutrino mass matrix $M_\nu$ invariant under this symmetry, {\it
i.e} $S_{1\nu}^T M_\nu S_{1\nu}=M_\nu$, can be written as
\be\label{mnu}
V_\nu^T M_\nu V_\nu =\left(\ba{ccc}
0&m&0\\
m&0&0\\
0&0&M\\
\ea \right)~,\ee
where $m$ and $M$ are complex parameters which are not fixed by the symmetry. The above $M_\nu$ is
diagonalized by $U_\nu = V_\nu R_{12}(\pi/4)Q$, where $R_{12}(\pi/4)$ represents a maximal
rotation in 1-2 plane and $Q={\rm Diag.}(i,1,1)$. The diagonal $M_\nu$ has two degenerate
eigenstates of mass $|m|$ to be identified as the solar pair and third eigenstate with mass $|M|$.
Two limits $|m| \ll |M|$ and $|m| \gg |M|$ correspond to the normal and the inverted ordering in
neutrino masses respectively. A special case of the latter with $|M|=0$ corresponds to an enlarged
residual symmetry which can be written as
\be \label{s2nu}
S_{2\nu}\equiv V_\nu~{\rm Diag.}(\eta,\eta^*,\beta)~V_\nu^\dagger ~\ee
with $\beta^k=1$ and $k\geq 3$. $M_\nu$ invariant under this
symmetry will describe a massless plus a Dirac state. In both these cases described above, the
lepton mixing matrix is given as 
\beqa \label{pmns}
U_{\rm PMNS} \equiv U &=& P_l~(V_l^\dagger V_\nu
R_{12}(\pi/4) Q)~R_{12}(\theta_x)P_{\beta_2} \equiv P_l U^0 R_{12}(\theta_x)P_{\beta_2}, \eeqa
where $P_l$ is a diagonal phase matrix with arbitrary phases and $V_l$ is defined in  Eq.
(\ref{talpha}). $R_{12}(\theta_x)$ represents an arbitrary rotation in 1-2 plane allowed due to the
degeneracy in neutrino masses and $P_{\beta_2}={\rm Diag.}(1,1,e^{i\beta_2/2})$ is a Majorana phase
matrix associated with only third neutrino. Both $\theta_x$ and $P_{\beta_2}$ cannot be fixed by the
symmetries $S_{1\nu}$ or $S_{2\nu}$ and $T_l$. However, the matrix $U^0$ is completely determined,
up to the freedom in interchanging the rows and the first two columns, once $S_{1\nu}$ or $S_{2\nu}$
and $T_l$ are known. In the following, we shall assume that
$S_{1\nu}$ or $S_{2\nu}$ along with $T_l$ represent the residual symmetries of a neutrino and the
charged lepton mass matrices and look for a symmetry groups $G_f$ which contain them. The choice of
$G_f$ and $S_{1\nu}$, $T_{l}$ within it will determine $V_\nu$, $V_l$ and hence $U^0$ in Eq.
(\ref{pmns}). A $T_l$ can be chosen such that ${\rm det}(T_l)=1$ and $G_f$ can be a DSG of
$SU(3)$ when $|m|,|M|\neq0$. In contrast, the group  embedding $S_{2\nu}$, $T_l$ will be a DSG of
$U(3)$ when $|M|=0$. We shall discuss these two cases separately in section \ref{SU3-ana} and
\ref{U3} respectively.

At this point, it is important to identify the physical observables in $U_{\rm PMNS}$ which do not
depend on the arbitrariness present in Eq. (\ref{pmns}) and can be predicted at zeroth order from
the underlying groups only.  The following combinations \cite{Hernandez:2013vya}  of the elements of
$U_{\rm PMNS}$ are
independent of the arbitrary parameters $\theta_x$ and $\beta_2$: 
\beqa \label{invariance}
|U_{\alpha 3}| &=& |U^0_{\alpha 3}|, \nonumber \\
{\cal I}_\alpha\equiv{\rm Im}(U^*_{\alpha 1}U_{\alpha 2}) &=& {\rm
Im}(U^{0*}_{\alpha 1}U^0_{\alpha
2})~, \eeqa
where $\alpha=e,\mu,\tau$ and repeated indices do not mean summation. Not
all the three
combinations in the second equation are independent as ${\cal I}_e+{\cal I}_\mu+{\cal I}_\tau=0$
follows from the unitarity of $U$. The first of Eq. (\ref{invariance}) determines the mixing angles
$\theta_{23}$ and $\theta_{13}$ without any ambiguity. The other two equations determine the
specific combinations of the solar angle $\theta_{12}$, Dirac CP phase $\delta$ and a Majorana phase
$\beta_1$. These combinations in the standard parametrization are given as
\beqa \label{correlations}
 c_{12} s_{12} \sin\frac{\beta_1}{2} &=& \frac{1}{c_{13}^2}{\cal I}_e, \nonumber \\
 c_{12}^2\sin\left(\delta-\frac{\beta_1}{2}\right)+s_{12}^2\sin\left(\delta+\frac{\beta_1}{2}\right)
&=& \frac{1}{s_{23}c_{23}s_{13}}\left({\cal I}_\mu -\frac{(s_{23}^2
s_{13}^2-c_{23}^2)}{c_{13}^2}{\cal I}_e\right), \eeqa
where $s_{ij} =\sin\theta_{ij}$ and $c_{ij}= \cos\theta_{ij}$. The quantities in the right side of
the above equations can be determined from the underlying symmetries. Note that $\theta_{12}$,
$\delta$ and $\beta_1$ are unphysical if the solar neutrino pair is exactly
degenerate. The small
perturbations required to lift the degeneracy then determines the values of these observables.
However, the above combinations between these observables are fixed at the leading order
solely by the symmetries. Thus they are expected to receive small corrections  just
like $\theta_{23}$ and $\theta_{13}$.
Considering this, we
provide the leading order predictions not only for $\theta_{23}$ and $\theta_{13}$ but also for the
correlations given in Eqs. (\ref{correlations}) in the following analysis.

\section{DSG of \texorpdfstring{$SU(3)$}{SU(3)} and neutrino mixing} 
\label{SU3-ana}
All the DSG of $SU(3)$ have been systematically classified in terms of their generators
\cite{Miller,Fairbairn:1964sga,Bovier:1980gc,Luhn:2007yr,Luhn:2007uq,Escobar:2008vc}. They
are listed and further studied in
\cite{Grimus:2010ak,Grimus:2011fk,Grimus:2013apa,Ludl:2009ft,Ludl:2010bj,
Zwicky:2009vt,Merle:2011vy}. In the following, we shall consider all the DSG of $SU(3)$ having
three dimensional IR and look for viability of these groups as a flavour
symmetry for a Dirac neutrino and work out the predictions of the mixing angles. Specifically we
shall look for the minimal DSG of $SU(3)$ containing groups generated by $S_{1\nu}$ and
$T_l$ of Eqs. (\ref{snu}) and (\ref{talpha}) as subgroups. The $T_l$ will be assumed
to have determinant $+1$ and all three different eigenvalues. Since all eigenvalues of $S_{1\nu}$
are also different, both $V_\nu$ and $V_l$ get uniquely determined apart from a diagonal phase
matrix. Then using Eqs. (\ref{upmns}) and (\ref{invariance}) one can determine leading order
predictions for $\theta_{13}$ and $\theta_{23}$ and the correlations listed in Eq.
(\ref{correlations}). Small perturbations generating the solar scale is not expected to change these
predictions in a big way and we shall work out these predictions analytically and numerically for
various groups. All the
 DSG of $SU(3)$ having three dimensional IR are classified in three main categories: type C, type D
and some other that  do not fall in either of these two categories. In this section, we present a
complete analytical study of the first two categories of the groups for their viability as a
symmetry of degenerate solar pair together with realistic mixing pattern. A numerical analysis for
the groups in all three categories will be presented in the next section.

\subsection{\texorpdfstring{$SU(3)$}{SU(3)} subgroups of type C: The group series
\texorpdfstring{$C(n,a,b)$}{C(n,a,b)}}
The groups of type C, also characterized by a series $C(n,a,b)$, are isomorphic to groups
generated from the following $3\times 3$ matrices \cite{Grimus:2013apa,Ludl:2010bj}: 
\be \label{cn}
\ba{cc}
F(n,a,b)=\left(\ba{ccc}
\eta^a&0&0\\
0&\eta^b&0\\
0&0&\eta^{-a-b}\\ \ea \right),~~
E=\left(\ba{ccc}
0&1&0\\
0&0&1\\
1&0&0\\ \ea \right)\\ \ea,\ee
where $\eta=e^{2 \pi i/n}$ and $a,b=0,1,2,..,n-1$. We shall first discuss a member $\Delta(3
n^2)\equiv C(n,0,1)$ of this series and then generalize the consideration to all the groups in the
series. $\Delta(3 n^2)$ is isomorphic to $(Z_n\times Z_n)\rtimes Z_3$. Detailed properties of these
group are studied in \cite{Luhn:2007uq,Ishimori:2010au}. All of its IR are
either three or one dimensional. $E$ and diagonal matrix $F(n,0,1)={\rm Diag.}(1,\eta,\eta^*)$
provide generators of a faithful 3-dimensional representation. Their multiple products therefore
generate the entire group whose elements can be labeled as:
\beqa
\label{elements3nsquare}
 W(n,a,b)&\equiv& W=\left(
\ba{ccc}
\eta^a&0&0\\
0&\eta^b&0\\
0&0&\eta^{-a-b}\\
\ea \right),~
R(n,a,b)\equiv R=\left(
\ba{ccc}
0&0&\eta^a\\
\eta^b&0&0\\
0&\eta^{-a-b}&0\\
\ea \right), \nonumber \\
V(n,a,b)&\equiv& V=\left(
\ba{ccc}
0&\eta^a&0\\
0&0&\eta^b\\
\eta^{-a-b}&0&0\\
\ea \right).\eeqa
Here $a,b=0,1,2,..,n-1$ and hence each of the non-zero entries of above three matrices take $n$
different values corresponding to $n^{\rm th}$ roots of unity. Since one of the row in each of the
matrices is determined from the other two due to their unit determinant, we get total $3n^2$
independent elements of the group. In the following, we shall first assume that leptonic doublets
are assigned to the 3-dimensional representation given in Eq. (\ref{elements3nsquare}) and show that
basic conclusion remains unchanged even if they are assigned to any other 3-dimensional
representations.

In order to obtain a Dirac state through $\Delta(3n^2)$, one needs to find appropriate element of
the group which can be used as $S_{1\nu}$ in Eq. (\ref{snu}) and $T_l$ in Eq. (\ref{talpha}). There
exists numerous choices for these. To see this, we note that all the non-diagonal elements $R$ and
$V$ of the group posses identical set of eigenvalues $(1,\omega,\omega^2)$ with
$\omega=e^{2 \pi i/3}$. Using this, we can easily enumerate all possible choices of the residual
symmetries. $S_{1\nu}$ can either be (a) a diagonal element in set $W$ such that one of the non-zero
entry is 1 and others are complex or (b) any of the matrices contained in $R$ or $V$. Likewise, 
$T_l$ can be (1) any element in $W$ with unequal eigenvalues or (2) any element in $R$ or $V$.
Choices of $S_{1\nu}$, $T_l$ automatically determine the diagonalizing matrices $V_\nu$, $V_l$ and
hence $U^0$ in Eq. (\ref{pmns}). Non-trivial mixing occurs only when at least one of the residual
symmetries is chosen to be either $R$ or $V$. Explicitly, $R$ is diagonalized by:
\be \label{vab}
V_R(n,a,b)= {\rm Diag.}(1,\eta^b,\eta^{-a})~U_\omega~~\ee
with 
\be \label{uw}
U_\omega=\frac{1}{\sqrt{3}}\left( \ba{ccc}
1&1&1\\
1&\omega^2&\omega\\
1&\omega&\omega^2\\
\ea \right).\ee
Similarly, $V$ is diagonalized by $V_{V}(n,a,b)\equiv
{\rm Diag.}(1,\eta^{-a},\eta^{-a-b})~U_\omega^*$.

Combinations of possibilities for the choice of residual symmetries listed above determine
all the mixing patterns that can be generated using the $\Delta(3n^2)$ groups. As already
mentioned, the choice (a) and (1) leads to $V_\nu=V_l=I$ and vanishing $\theta_{13}$ and
$\theta_{23}$ in Eq. (\ref{pmns}). The choices, (a) and (2) or (b) and (1) contains either $V_\nu$
or $V_l$ proportional to identity while the other is given by $V_{R}(n,a,b)$ or $V_{V}(n,a,b)$. It
follows then from Eq. (\ref{pmns}) that $|U_{\alpha3}|$ has a democratic structure and implies
$\sin^2\theta_{13}=\frac{1}{3}$ which is very large. One therefore is left to consider the case (b)
and (2). There are four possible structures for $|U_{\rm PMNS}|$ in this case but all give
equivalent mixing and we explicitly consider the choice:
\be
\ba{cc}
S_{1\nu}=\left(
\ba{ccc}
0&0&\eta^{a_\nu}\\
\eta^{b_\nu}&0&0\\
0&\eta^{-a_{\nu}-b_{\nu}}&0\\
\ea \right),~~~
T_l=\left(
\ba{ccc}
0&0&\eta^{a_l}\\
\eta^{b_l}&0&0\\
0&\eta^{-a_{l}-b_{l}}&0\\
\ea \right)~,\ea \ee
The matrices diagonalizing above are respectively given by $V_\nu=V_{R}(n,a_\nu,b_\nu)P_{13}$ and
$V_l=V_{R}(n,a_l,b_l)$. The columns of the matrix $V_{R}(n,a,b)$ given in Eq. (\ref{vab}) correspond
to eigenvectors with eigenvalues $(1,\omega,\omega^2)$. If $S_{1\nu}$ is to define a Dirac pair with
first two degenerate eigenvalues then the first two columns of the diagonalizing matrix should
correspond to the eigenvalues which are complex conjugate to each other. This is done by inserting a
matrix  $P_{13}$ which interchanges the first and the third column of $V_R(n,a_\nu,b_\nu)$ to give
proper $V_\nu$.

\begin{figure}[!ht]
\centering
\includegraphics[width=0.5\textwidth]{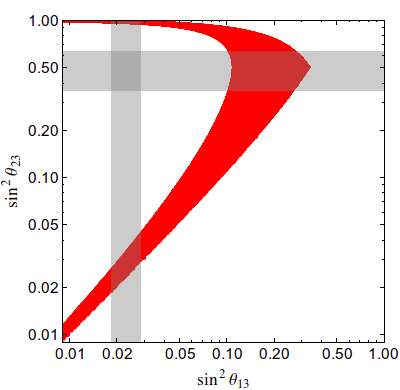}
\caption{Predictions for $\sin^2\theta_{13}$ and $\sin^2\theta_{23}$ obtained in case of the
C-type DSG of $SU(3)$. The allowed values lie at some discrete points in the red shaded region.
The horizontal and vertical gray bands correspond to the respective 3$\sigma$ ranges from
the global fits \cite{Capozzi:2013csa}.}
\label{fig1}
\end{figure}

$V_\nu$ and $V_l$ defined above determine the third column $|U_{\alpha 3}|$ of the mixing matrix
uniquely, see Eq. (\ref{invariance}). Explicitly, 
\beqa \label{ui3sq}
|U_{\alpha 3}|^2 &=&\frac{1}{9}\left(3+2 \cos x_\alpha+2\cos
y_\alpha+2\cos(x_\alpha+y_\alpha)\right)~.\eeqa
Here $\alpha=e,\mu,\tau$, and 
$$ x_\alpha=\frac{2 \pi}{n}\left(b_\nu-b_l+\frac{n}{3}
p_\alpha\right),~~~y_\alpha=\frac{2 \pi}{n}\left(a_\nu-a_l+\frac{n}{3} p_\alpha\right)$$
with $p_\alpha=(0,1,2)$. The entries  $|U_{\alpha 3}|^2$ can be interchanged by reordering the
eigenvalues of $T_l$. Using this freedom, we define
\be \label{s13}
\sin^2\theta_{13}={\rm Min}(|U_{\alpha 3}|^2),~~ \sin^2\theta_{23}~{\rm or~}
\cos^2\theta_{23}={\rm Max}(|U_{\alpha 3}|^2)~.\ee
Note that $a_\nu,a_l,b_\nu,b_l$ vary over all possible $n^{\rm th}$ roots of unity corresponding to
different choices of $S_{1\nu}$ and $T_l$ among the elements of $\Delta(3 n^2)$ and varying them
over all roots exhaust all predictions within the group. It is easy to show that there does
not exist any values of these parameters which can simultaneously give correct $\theta_{13}$ and
$\theta_{23}$. We demonstrate this graphically. Let us treat $x_\alpha$ and $y_\alpha$ as continuous
variables and vary them in the entire range $0$-$2\pi$. The allowed discrete choices for
$a_{\nu,l},b_{\nu,l}$ and $n$ will clearly span a subset of this continuous range.
Following this way and using Eqs.(\ref{s13}), the predictions for $\theta_{13}$ and
$\theta_{23}$ are depicted in Fig. \ref{fig1}.
All the possible predictions for $\theta_{13}$ and $\theta_{23}$ for the group series $\Delta(3 n^2)$
lie as discrete points in the red shaded region. As can be seen from figure, $\Delta(3 n^2)$
groups can at most give good leading order prediction either for $\theta_{13}$ or
$\theta_{23}$ but not for both. If one is near to the 3$\sigma$ limit then the other is far
away from it. Thus these groups are not suitable for the descriptions of a Dirac neutrino with
correct leading order predictions of mixing angles.

In deriving the above conclusion, we had used the fact that three generations of leptonic doublets
transform as a triplet representation generated using $E$ and $F(n,0,1)$. The group admits other
3-dimensional IR labeled by $3_{(k,l)}$ \cite{Luhn:2007uq}. These are
generated using $E, F(n,k,l)$ given in Eq. (\ref{cn}) with the provision that values of $(k,l)$
giving equivalent representation should be excluded \cite{Luhn:2007uq}. The forgoing argument gets
carried over even when leptons are assigned to any general triplet IR $3_{(k,l)}$ and not just
$3_{(0,1)}$ since it only used the basic texture for elements of the $\Delta(3 n^2)$ groups given in
Eq. (\ref{elements3nsquare}). The same texture gets carried over for all the triplet representation
except that non-zero entries in these elements now need not span all the $n^{th}$ roots. Thus the
leading order prediction will only be a subset of the ones predicted using the representation
$3_{(0,1)}$. Since the latter  fails in simultaneously giving correct mixing angles, assigning
leptons to a different triplet will not alter the general conclusion.

The argument in the forgoing para can be used not just to rule out other IR of $\Delta(3 n^2)$ but
the entire group series $C(n,a,b)$. This follows from the observation \cite{Ludl:2010bj} that the
generators of the series $C(n,a,b)$ as given in Eq. (\ref{cn}) also represent 3-dimensional IR of
the groups $\Delta(3 n^2)$. These groups are isomorphic to matrix groups generated using the IR
of $\Delta(3 n^2)$ and thus any of the groups $C(n,a,b)$ cannot lead to correct $\theta_{13}$ and
$\theta_{23}$ simultaneously. This negative result is useful in discarding large number of small DSG
of $SU(3)$ from our consideration as a possible candidate of symmetry in the lepton sector. For
example, among the first 59 DSG of $SU(3)$ of order $<512$ listed in \cite{Ludl:2010bj}, 46 groups
fall in this category and can be ruled out using the above results.

\subsection{\texorpdfstring{$SU(3)$}{SU(3)} subgroups of type D: The group series
\texorpdfstring{$D(n,a,b;d,r,s)$}{D(n,a,b;d,r,s)}}
The groups characterized by a series $D(n,a,b;d,r,s)$ are the groups obtained by adding the
following matrices to the generators $E$ and $F(n,a,b)$ of the series $C(n,a,b)$
\cite{Escobar:2008vc,Grimus:2013apa}:
\be
\label{g}
G(r,d,s)=\left(
\ba{ccc}
\delta^r&0&0\\
0&0&\delta^s\\
0&-\delta^{-r-s}&0\\
\ea \right)~,\ee
where $\delta=e^{2\pi i/d}$ with integer $d$ and $r,s=0,1,...,d-1$. We shall be using the following 
important results in our discussion of these groups:
\begin{enumerate}[(1)]
\item $\Delta(6 n^2)\equiv D(n,0,1;2,1,1)$ provides one of the infinite series within the type D
groups.
\item Every $D(n,a,b; d,r,s)$ is a subgroup of the group $\Delta(6 g^2)$ with $g={\rm lcm}~(n,d,2)$
\cite{Zwicky:2009vt}.
\item All the D-type groups are contained in only one of the three infinite series of groups
\cite{Grimus:2013apa}, namely $\Delta(6 n^2)$, $Z_3\times \Delta(6 n^2)$ and $D^1_{9n,3n}$ which is
isomorphic to $(Z_{9n}\times Z_{3n})\rtimes S_3$.
\end{enumerate}

Let us first consider the $\Delta(6 n^2)=D(n,0,1;2,1,1)$ groups whose finding can be used to make
some general statement on the other
groups in the series. Earlier study of these groups in the context of three non-degenerate neutrinos
was presented in \cite{King:2013vna}.  Addition of the generator $G(2,1,1)$ lead to
$3n^2$ new elements apart from the $3n^2$ elements, Eq. (\ref{elements3nsquare}), of $\Delta(3
n^2)$:
\beqa
\label{elements6nsquare}
S&\equiv& S(n,a,b)=-\left(
\ba{ccc}
\eta^a&0&0\\
0&0&\eta^b\\
0&\eta^{-a-b}&0\\
\ea \right),~~
T\equiv T(n,a,b)=-\left(
\ba{ccc}
0&0&\eta^b\\
0&\eta^a&0\\
\eta^{-a-b}&0&0\\
\ea \right)~,\nonumber\\
U&\equiv& U(n,a,b)=-\left(
\ba{ccc}
0&\eta^b&0\\
\eta^{-a-b} &0&0\\
0&0&\eta^a\\
\ea \right).\eeqa
Here also $a,b=0,1,..,n-1$. These new elements give more freedom in choosing residual symmetries
compared to $\Delta(3 n^2)$. To enumerate these choices, let us note that eigenvalues of $S,T,U$ are
given by $(-\eta^a,\eta^{-\frac{a}{2}},-\eta^{-\frac{a}{2}}$). The corresponding diagonalizing
matrices are:
\beqa \label{stu1}
V_{S}(a,b)&=&\frac{1}{\sqrt{2}}\left(
\ba{ccc}
\sqrt{2}&0&0\\
0&1&\eta^{b+a/2}\\
0&-\eta^{-b-a/2}&1\\
\ea \right),~~
V_{T}(a,b)=\frac{1}{\sqrt{2}}\left(
\ba{ccc}
1&0&\eta^{b+a/2}\\
0&\sqrt{2}&0\\
-\eta^{-b-a/2}&0&1\\
\ea \right),\nonumber \\
V_{U}(a,b)&=&\frac{1}{\sqrt{2}}\left(
\ba{ccc}
1&\eta^{b+a/2}&0\\
-\eta^{-b-a/2}&1&0\\
0&0&\sqrt{2}\\
\ea \right).\eeqa
The $S,T,U$ contain two complex conjugate  eigenvalues 
only for even $n$ and when $a=n/2$. The corresponding eigenvalues are $(i,-i,1)$. Except in this
special case, the above $S,T,U$ cannot serve as the desired residual symmetry of neutrino mass
matrix. It follows from the structures of the diagonalizing matrices and Eq. (\ref{pmns}) that if
$S_{1\nu}$ or $T_l$ is diagonal, {\it i.e.} $\sim W$, then the third column of $U_{\rm PMNS}$ is
either
$1/\sqrt{3}(1,1,1)^T$ or $1/\sqrt{2}(0,1,1)^T$ or its permutations leading to very large or
vanishing $\theta_{13}$. The latter result is similar to the prediction of the \mt symmetry. Thus
both $S_{1\nu}$ and $T_l$ must be non-diagonal in order to obtain non-trivial prediction. Thus, we
are left with four
cases:
\begin{enumerate}[(i)]
\item  Both $S_{1\nu},T_{l}\in \{R,V\}$ 
\item $T_l\in \{R,V\}$ and
$S_{1\nu}\in\{S,T,U\}$ with eigenvalues $(i,-i,1)$.
\item Both $S_{1\nu},T_{l}\in \{S,T,U\}$ and
$S_{1\nu}$ has eigenvalues $(i,-i,1)$
\item $S_{1\nu}\in \{R,V\}$ and $T_l\in\{S,T,U\}$
\end{enumerate}
The case (i) has already been discussed and leads to the results shown in Fig. \ref{fig1}. The
cases (ii) and (iii) respectively fix the third column of $U_{\rm PMNS}$ to $1/\sqrt{3}(1,1,1)^T$
and $1/\sqrt{2}(0,1,1)^T$ or its permutations. The case (iv) can give non-zero and small
$\theta_{13}$ and we consider it explicitly. Let us choose $S_{1\nu}=R(n,a_\nu,b_\nu)$ and
$T_l=T(n,a_l,b_l)$. Other choices within this set do not give any new results. The corresponding
diagonalizing matrices are given by $V_\nu=V_{R}(n,a_\nu,b_\nu)P_{13}$ and $V_l=V_{T}(n,a_l,b_l)$
with $V_{R}$ and $V_{T}$ as in Eqs. (\ref{vab}) and (\ref{stu1}) respectively. As before, a
permutation matrix $P_{13}$ is introduced to ensure that the eigenvectors are ordered according to
the eigenvalues $(\omega^2,\omega,1)$. The third column of the mixing matrix then follows from
Eq. (\ref{pmns}) and leads to the following predictions:
\beqa\label{6nprediction}
\sin^2\theta_{13}&=&\frac{2}{3}{~\rm
Min}\left(\cos^2\left(\pi \frac{b_l-a_\nu+\frac{a_l}{2}}{n}\right),~\sin^2\left(\pi
\frac{b_l-a_\nu+\frac{a_l}{2}}{n} \right)\right)~, \nonumber\\
(\sin^2\theta_{23}~{\rm or~} \cos^2\theta_{23})&=&\frac{1}{3\cos^2\theta_{13}}~.\eeqa

One could rearrange the entries in the third column by changing the order of the eigenvalues of
$T_l$. This corresponds to interchange $\sin^2\theta_{23}\leftrightarrow \cos^2\theta_{23}$ and one
gets two possibilities given in the last equation above. Note that the second in Eq.
(\ref{6nprediction}) is universal and is dictated by the textures of the chosen $S_{1\nu}$, $T_l$
and not by the actual values of the parameters $a_{\nu},b_l,a_l$. This means that this 
prediction can be obtained in all the groups of the series
$\Delta(6n^2)$ by choosing $S_{1\nu}$ and $T_l$ as in this case. One may regard this universal
prediction as a very good zeroth order choice. Inserting the best fit value
$\sin^2\theta_{13}=0.023$ in Eq. (\ref{6nprediction}) gives $\sin^2\theta_{23}=0.341~{\rm or~}
0.658$ which is though outside the allowed 3$\sigma$ range is quite close to it and small
perturbation may bring it within the required range. The other invariant quantities expressed in Eq.
(\ref{invariance}) can also be determined for the above choice of $S_{1\nu}$ and $T_l$ as
\beqa \label{invariance-Dtype}
{\cal I}_\mu = 0,~{\cal I}_e = -{\cal I}_\tau &=&
\pm\frac{1}{2\sqrt{3}}\sin\left(
2\pi\frac{b_l-a_\nu+\frac{a_l}{2}}{n}\right)\nonumber \\
&=&\pm \frac{\sqrt{3}}{2}\sin\theta_{13}\cos\theta_{13}\cos\theta_{23}~~~~{\rm
for}~\sin^2\theta_{23}\cos^2\theta_{13}=\frac{1}{3},\nonumber \\
{\cal I}_\tau = 0,~{\cal I}_e = -{\cal I}_\mu &=&
\pm\frac{1}{2\sqrt{3}}\sin\left(
2\pi\frac{b_l-a_\nu+\frac{a_l}{2}}{n}\right)~\nonumber \\
&=&\pm\frac{\sqrt{3}}{2}\sin\theta_{13}\cos\theta_{13}\sin\theta_{23} ~~~~{\rm
for}~\cos^2\theta_{23}\cos^2\theta_{13}=\frac{1}{3}, \eeqa
where we have used Eq. (\ref{6nprediction}) in writing the above equation.
Note that at least one of the ${\cal I}_\alpha$s is vanishing for all the groups of the series
$\Delta(6n^2)$. In addition to the above, there are two other universal predictions possible within
the groups in the $\Delta(6n^2)$ series. First one is obtained when $a_l,b_l,a_\nu$ are chosen as
zero which is always an allowed choice. From Eq. (\ref{6nprediction}), one gets in this case
$\theta_{13}=0$ and $\sin^2\theta_{23}=\frac{1}{3}$. The other prediction corresponds to the case
(iii) discussed above, {\it i.e.} $\theta_{13}=0$ and $\theta_{23}=\frac{\pi}{4}$, and follows in
all the $\Delta(6n^2)$ groups with even $n$.

\begin{figure}[!ht]
\centering
\includegraphics[width=0.5\textwidth]{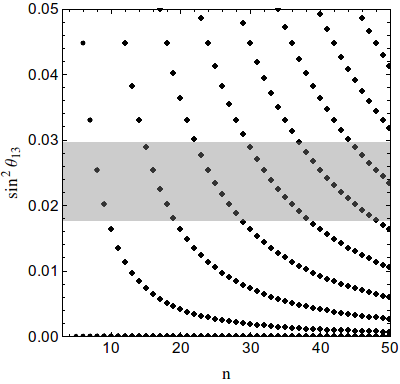}
\caption{Predictions for $\sin^2\theta_{13}$ as function of $n$ for $n\leq 50$ in case of the
$\Delta(6n^2)$ groups. The horizontal gray band corresponds to the 3$\sigma$ region in
$\sin^2\theta_{13}$ allowed by the global fits \cite{Capozzi:2013csa}.}
\label{fig2}
\end{figure}

There exists several choices for $n$ and $a_\nu,b_\nu,a_l,b_l$ which can give correct $\theta_{13}$
in the 3$\sigma$ range allowed by the global fits \cite{Capozzi:2013csa}. These are group specific.
We explore this numerically for first few values of $n \le 50$. The results are displayed in Fig.
\ref{fig2}. It is seen from the figure that the smallest such value occurs for $n=8$ which
corresponds to an order 324 group $\Delta(6\cdot8^2)$. Further, for the values of $8< n\le 50$, all
$n$  except $10\le n \le 14$ and $n=20,21$ can lead to the value of $\theta_{13}$ preferred by
global fits at 3$\sigma$.

The above considerations can be easily generalized to all the DSG of $SU(3)$ of type D since latter
are always subgroups of some $\Delta(6g^2)$ as already mentioned. As a consequence, all the elements
of any D-type groups have the same six possible textures as in the case of $\Delta(6 n^2)$, Eqs.
(\ref{elements3nsquare}) and (\ref{elements6nsquare}). In particular, choices of $S_{1\nu}$ and
$T_l$ as discussed in case (iv) for $\Delta(6 n^2)$ always exist for any type D groups and the
resulting prediction, Eq. (\ref{6nprediction}) always holds but allowed values of parameters
$a_l,b_l,a_\nu$ may now form a subset of the possible values in case of $\Delta(6g^2)$. In
particular, the universal prediction, second of Eqs. (\ref{6nprediction}) which is independent of
the values of these parameters will always follow in all the groups of type D. Allowed values of
$\theta_{13}$ will be a subset of the prediction in Fig \ref{fig2}. We shall explore this
numerically for the group series $D^1_{9n,3n}$ in the next subsection along with the other groups.

\section{Numerical study of DSG of \texorpdfstring{$SU(3)$}{SU(3)}}
\label{SU3-num}
We now numerically study various DSG of $SU(3)$. This study includes (1) all DSG of $SU(3)$ with
order $<512$ as tabulated by Ludl in \cite{Ludl:2010bj}, (2) the remaining groups $A_5$,
$\Sigma(108)$, $\Sigma(168)$, $\Sigma(216)$, $\Sigma(648)$, $\Sigma(1060)$ which are not of
type C or D and (3) some of the larger groups of type D labeled as $D^1_{m,n}$ by Grimus and
Ludl \cite{Grimus:2013apa}. Numerical procedure followed is the following. We first construct all
the elements of a group falling in the three categories described above from the defining
generators. We then identify all possible elements which can be chosen as possible residual
symmetries $S_{1\nu}$ and $T_l$ of the neutrino and the charged leptons respectively such that
$S_{1\nu}$ and $T_l$ do not commute and construct all possible mixing matrices from their
diagonalization.
\begin{table}[!ht]
\begin{small}
\begin{center}
\begin{tabular}{cccccc}
 \hline
 \hline
  Group ~&~ GAP code~&~$\sin^2\theta_{13}$ ~&~
  $\sin^2\theta_{23}$~&~${\cal I}_e$=Im($U^*_{e1}U_{e2}$)~&~${\cal I}_\mu$=Im($U^*_{\mu
1}U_{\mu 2}$)\\
  \hline
  $\Delta(6\cdot6^2)$ & \group{216,95} &  0.045 & 0.349 & $\pm 0.144$ & 0\\
  $\Delta(6\cdot7^2)$ & \group{294,7}  &  0.033  & 0.345 & $\pm 0.125$ & 0\\
  $\Delta(6\cdot8^2)$ & \group{384,568}&  0.025$^\bigstar$  & 0.342 & $\pm 0.11$ & 0 \\
  $\Delta(6\cdot9^2)$ & \group{486,61} &  0.02$^\star$   & 0.34 & $\pm 0.099$ & 0\\
  \hline
  $D_{9,3}^1\cong (Z_9\times Z_3)\rtimes S_3$ & \group{162,14}&  0.02$^\star$ & 0.34 & $\pm
  0.099$ & 0 \\
  $D_{18,6}^1\cong (Z_{18}\times Z_6)\rtimes S_3$ & \group{648,259}& 0.005 & 0.335 & $\pm 0.05$ &
0\\ 
                  &              &  0.02$^\star$ & 0.34 & $\pm 0.099$ & 0 \\ 
                  &              &  0.045 & 0.349 & $\pm 0.144$ & 0\\
  $D_{27,9}^1\cong (Z_{27}\times Z_9)\rtimes S_3$ & \group{1458,658}& 0.002 & 0.334 & $\pm 0.034$ &
0
\\ 
                  &              &  0.009 & 0.336 & $\pm 0.067$ & 0 \\ 
                  &              &  0.02$^\star$ & 0.34 & $\pm 0.099$ & 0 \\ 
                  &              &  0.035 & 0.346 & $\pm 0.13$ & 0\\  
 \hline
                  $A_5\cong \Sigma(60)$ & \group{60,5}&  0.035 & 0.5$^\star$ & 0 & $\pm 0.094$ \\
  $PSL(2,7)\cong \Sigma(168)$ & \group{168,42}& 0.023$^\bigstar$ & 0.455$^\bigstar$ & $\pm 0.377$ &
$\mp0.255$\\
                               &               & 0.027$^\star$ & 0.293 & $\pm 0.333$ & $\mp
0.185$ \\
                               &               & 0.044 & 0.262 & $\pm0.353$ & $\mp0.331$ \\ 
  $\Sigma(648)$ & \group{648,532}& 0.012 & 0.279 &$\pm 0.342$ & $\mp0.284$\\
                &               & 0.019$^\star$ & 0.341 &$\pm 0.406$ &
$\mp0.307$\\
                &               & 0.02$^\star$  & 0.34  &$\pm 0.099$ & 0\\
  $\Sigma(1080)$ & \group{1080,260}& 0.006 & 0.364$^\star$ & $\pm0.392$ & $\mp0.228$\\
                &               & 0.016 & 0.226 &$\pm0.27$ & $\mp0.167$\\
                &               & 0.035 & 0.458$^\bigstar$ & $\pm0.283$ & $\mp 0.079$\\
                &               & 0.035 & 0.5$^\star$ & 0 & $\pm 0.094$ \\
                &               & 0.039 & 0.319 & $\pm 0.153$& $\mp0.019$\\
                &               & 0.042 & 0.392$^\bigstar$ & $\pm 0.436$& $\mp 0.27$\\
                &               & 0.045 & 0.349 & $\pm 0.144$& 0\\
\hline
\hline
\end{tabular}
\end{center}
\end{small}
\caption{Predictions for $\theta_{23}$, $\theta_{13}$ and ${\cal I}_\alpha$ with two degenerate
neutrinos and third massive neutrino from a scan of various DSG of $SU(3)$. The prediction for
$\sin^2\theta_{23}$ is chosen to be in the first octant. Only those groups which give
$0<\sin^2\theta_{13}\leq 0.05$ and $0.2\leq\sin^2\theta_{23}\leq 0.5$ are included. The predictions
marked by $^\bigstar$ $(^\star)$ are within the $1\sigma$ (3$\sigma$) interval of the global fits
\cite{Capozzi:2013csa}. The group \group{$g,j$} refers to the $j^{\rm th}$ finite group of order $g$
as classified in the Small Group Library in GAP \cite{GAP4}.}
\label{SU3-results}
\end{table}
The results are displayed in Table \ref{SU3-results} in which we collect only the solutions which
give $0<\sin^2\theta_{13}\le0.05$ and $0.2<\sin^2\theta_{23}\le0.5$. We have chosen $\theta_{23}$ in
the first octant and its other predicted value can be obtained by reading fourth column in Table
\ref{SU3-results} as $\cos^2\theta_{23}$. We also give the predictions for ${\cal I}_e$ and ${\cal
I}_\mu$ which determine certain correlations between the solar angle, Dirac CP phase and one of the
Majorana phases as displayed in Eq. (\ref{correlations}).
Numerical results are consistent with the analytic results discussed in the last section. Noteworthy
features of the table are:
\begin{itemize}
\item None of the C-type groups feature in the table as they cannot give correct $\theta_{13}$
and $\theta_{23}$ simultaneously.
\item The D-type groups shown in table are first few successful members of the
series $D^0_{n,n}=\Delta(6n^2)$ and $D^1_{9n,3n}$. In all the cases, the atmospheric mixing angle
determined numerically satisfies the second of Eq. (\ref{6nprediction}). Universal prediction
$\theta_{13}=0$ and $\sin^2\theta_{23}=\frac{1}{3}$ are found to be true for all the D-type groups
for $n\ge3$ and we do not show them in the Table. Likewise $\Delta(6 n^2)$ with even $n$ and $n\ge4$
also give the \mt symmetric prediction. These cases will need significant non-leading corrections to
explain the experimental value. The last two are the only possible prediction for $n \le 5$. Because
of this, the familiar small groups in type D like $S_4\cong \Delta(24)$, $\Delta(96)$ do not
appear in Table \ref{SU3-results}. Various groups with $n\geq8$ can predict $\theta_{13}$ within the
allowed
3$\sigma$ range.
\item We also show prediction obtained in case of three other groups namely $D^1_{9,3}$,
$D^1_{18,6}$ and $D^1_{27,9}$ with order 162, 648 and 1458 respectively. They do not lead to any new
useful leading order prediction for $\theta_{13}$. 
\item The prediction for the atmospheric mixing angle in case of the non D-type groups are
better. In particular, the group $\Sigma(168)\cong PSL(2,7)$ predicts $\theta_{13}$ very close to
the best fit value and $\theta_{23}$ within 1$\sigma$
of the best fit value. The mixing angle predictions for this group in the
case of three non degenerate neutrinos were presented in \cite{deAdelhartToorop:2011re}.
Our predictions are different due to difference in the chosen residual symmetry.
\item The $A_5\cong \Sigma(60)$ is the only group in the Table \ref{SU3-results} which belongs to
the category of finite von-Dyck groups \cite{Hernandez:2012ra,Hernandez:2012sk} investigated earlier
as a symmetry of degenerate solar pair in \cite{Hernandez:2013vya}. Our results of this group are in
agreement with their findings. In particular, this group predicts vanishing $\beta_1$ and maximal
$\delta$ from Eq. (\ref{correlations}) and Table  \ref{SU3-results} if solar angle is non-zero as
would be the case with any realistic perturbation. We shall explore this prediction further in
section \ref{examples}.
\end{itemize}

\section{Flavour symmetries for degenerate neutrino pair and a massless neutrino}
\label{U3}
We now discuss a different class of symmetries which provide a good zeroth order spectrum for an
inverted hierarchy, {\it i.e.} two degenerate neutrinos $(\nu_1,\nu_2)$ which are mix of flavour
states and a third massless neutrino $\nu_3$.

Flavour symmetries for a massless neutrino are extensively discussed in
\cite{Joshipura:2013pga,Joshipura:2014pqa} in the case when other two neutrinos are non-degenerate.
We now extend the same analysis to the case when massive neutrinos are degenerate. As discussed,
this can be obtained if the residual symmetry of the neutrino masses is given by Eq. (\ref{s2nu}).
Thus, we look for groups $G_f$ which contain groups generated by $S_{2\nu}$ and $T_l$ as sub-groups.
$G_f$ has to be a subgroup of $U(3)$ since ${\rm det}(S_{2\nu})\neq \pm1$. Unlike in the case of
$SU(3)$,
not all the DSG of $U(3)$ are classified. But Ludl \cite{Ludl:2010bj} has identified all the DSG of
$U(3)$ with order $<512$ and has also found several infinite series of groups (see also,
\cite{Parattu:2010cy}). In the following, we numerically investigate all the groups with order
$<512$ having three dimensional IR as tabulated by Ludl. Before doing this, we
discuss some general analytic properties useful in understanding numerical results to be derived.

We divide DSG of $U(3)$ in two classes  called X and Y in \cite{Joshipura:2014pqa}. The groups in
class X are generated by combination of six matrices having the same textures as $R,S,T,U,V,W$
defined earlier in Eqs. (\ref{elements3nsquare},\ref{elements6nsquare}). But now determinant of
theses generators is not required to be one and thus each of these are functions of four parameters
$\eta,a,b,c$ instead of three. Explicit forms of these are given by Ludl \cite{Ludl:2010bj}. Given
these forms, one argues \cite{Joshipura:2014pqa} that all the elements of any group $G_f$ in
category X has six possible textures labeled as $\tilde{R}, \tilde{S}, \tilde{T}, \tilde{U},
\tilde{V},\tilde{W}$. These are obtained respectively from $R,S,T,U,V,W$ by replacing non-zero
entries by some root of unity. For example,
\be \label{tilder}
\tilde{R}(\eta_1,\eta_2,\eta_3)=\left(
\ba{ccc}
0&0&\eta_1\\
\eta_2&0&0\\
0&\eta_3&0\\
\ea
\right),~\tilde{S}(\eta_1,\eta_2,\eta_3)=\left(
\ba{ccc}
\eta_1&0&0\\
0&0&\eta_2\\
0&\eta_3&0\\
\ea \right),\ee
where $\eta_{1,2,3}$ are some roots of unity. The eigenvalues
of $\tilde{A}\in\{\tilde{R},\tilde{V}\}$ are given by ${\rm
det}(\tilde{A})^{\frac{1}{3}}(1,\omega,\omega^2)$ with $\omega^3=1$. Thus they do not contain a
pair of eigenvalues which are complex conjugate to each other as long as ${\rm det}(\tilde{A}) \neq
1$. Such elements cannot therefore be chosen as a possible $S_{2\nu}$. The latter can come either
from the diagonal structure $\tilde{W}$ or non-diagonal structure $\tilde{S},\tilde{T},\tilde{U}$
having only one non-zero diagonal entry say $\eta_1$. The other non-zero entries are labeled as
$\eta_2,\eta_3$.
 The eigenvalues of matrices in this class are
$\eta_1,\pm\sqrt{\eta_2\eta_3}$ and they can have a complex pair of eigenvalues which can be used to
obtain a massless state. The diagonal matrix $\tilde{W}$ can also be chosen as either $T_l$ or
$S_{2\nu}$ for appropriate values of its elements. But choosing either $S_{2\nu}$ or $T_l$ as
diagonal does not give correct mixing pattern. Thus, one has essentially two non-trivial
possibilities: (i) $T_l \in \{\tilde{R},\tilde{V}\}$ and $S_{2\nu} \in
\{\tilde{S},\tilde{T},\tilde{U}\}$ and (ii) both $T_l$ and $S_{2\nu} \in
\{\tilde{S},\tilde{T},\tilde{U}\}.$

Since scenario under consideration applies only to the inverted hierarchy, the flavour composition
of the massless state determines the third column of the mixing matrix and hence $\theta_{13}$ and
$\theta_{23}$. Clearly, the case (ii) can only give \mt symmetric structure as a non-trivial
prediction. Let us thus consider the case (i) above and choose as an example, $S_{2\nu}=\tilde{S}
(\eta_{1\nu},\eta_{2\nu},\eta_{3\nu})$ and $T_l=\tilde{R}(\eta_{1l},\eta_{2l},\eta_{3l})$ explicitly
given in Eq. (\ref{tilder}). Two eigenvalues of $S_{2\nu}$ can be complex conjugate if\footnote{The
alternative possibility, $\eta_1\not=\pm 1,\eta_2\eta_3=-1$ gives a massless state with the \mt
symmetric structure as already found in \cite{Joshipura:2014pqa}.}
\be \label{pm}
\eta_{1\nu}^*=\pm (\eta_{2\nu}\eta_{3\nu})^{\frac{1}{2}}~.\ee
Eigenvector of $S_{2\nu}$ corresponding to the massless state in this case
is given by $$|\psi_0\rangle
=\frac{1}{\sqrt{2}}(0,1,\mp (\eta_{2\nu}^*\eta_{3\nu})^{\frac{1}{2}})^T$$
The $U_l$ diagonalizing $T_l$ is given by 
$$U_l=V_{\tilde{R}}={\rm Diag.}(1,p_l^*\eta_{2l},p_l\eta_{1l}^*)U_{\omega}~,$$
where $U_{\omega}$ is defined in Eq. (\ref{uw}) and
$p_l=(\eta_{1l}\eta_{2l}\eta_{3l})^{\frac{1}{3}}$. One thus gets flavour
structure of massless state by operating $U_l$ on $|\psi_0\rangle$. This leads to 
\be \label{ui3sqm}
|U_{\alpha 3}|^2=\frac{1}{6}|p_l\eta_{2l}^*\lambda_\alpha\mp
p_l^*\lambda_\alpha^*\eta_{1l}(\eta_{2\nu}^*\eta_{3\nu})^{\frac{1}{2}}|^2~,\ee
where $\alpha=e,\mu,\tau$ and $\lambda_\alpha=(1,\omega,\omega^2)$.

This is to be compared with result derived in \cite{Joshipura:2014pqa} where structure of massless
state for all the groups in category X was determined. In particular, it was shown that only
possible flavour probabilities for a massless state in case of the inverted hierarchy is
$(0,1/2,1/2)$ and its permutation. Unlike this, one can now find non-trivial $\theta_{13}$ with the
inverted hierarchy when the solar pair is degenerate at the leading order. The reason for this can
be traced to different structure of $S_{2\nu}$  which implies a different structure for
$|\psi_0\rangle$ compared to one with non-degenerate solar pair at the leading order.
\begin{table}[!ht]
\begin{small}
\begin{center}
\begin{tabular}{cccccc}
 \hline
 \hline
  GAP code ~&~ Group ~&~ $\sin^2\theta_{13}$ ~&~ $\sin^2\theta_{23}$~&~${\cal
I}_e$=Im($U^*_{e1}U_{e2}$)~&~${\cal I}_\mu$=Im($U^*_{\mu 1}U_{\mu 2}$)\\
  \hline
  \group{96,65} & $S_4(3)$ & 0.045 & 0.349 & $\pm 0.144$ & 0\\
  \group{324,13} &   &  0.02$^\star$ & 0.399$^\bigstar$ & $\pm0.157$ & $\mp0.029$\\
  \group{384,571} & $\Delta(6\cdot4^2,3)$ & 0.045 & 0.349 & $\pm 0.144$ & 0\\
  \hline
  \group{216,25} &   &  0.033 & 0.388$^\star$ & $\pm0.108$ & $\mp0.154$\\
  \group{432,273} &   &  0.033 & 0.388$^\star$ & $\pm0.108$ & $\mp0.154$\\
\hline
\hline
\end{tabular}
\end{center}
\end{small}
\caption{Results obtained for DSG of $U(3)$ of order $<512$ as listed in \cite{Ludl:2010bj}. The
other details are same as in the caption of Table \ref{SU3-results}.}
\label{U3-results}
\end{table}
Eq. (\ref{ui3sqm}) leads to non-trivial prediction for the third column which we now explore
numerically for all the 75 groups tabulated as DSG of $U(3)$ in \cite{Ludl:2010bj}. Inspection of
Table V in \cite{Ludl:2010bj} coupled with analytic argument shows that only 22 of the 75
groups in category
X can have element in class (i) and thus can give non-trivial prediction \cite{Joshipura:2014pqa}.
Along with these, five
groups in category Y also have to be studied numerically. We generate all the elements of these 27
groups numerically. Then identify all the elements with a pair of complex conjugate eigenvalues and
take them as possible $S_{2\nu}$ and elements with three distinct eigenvalues are chosen as $T_l$.
By diagonalizing these sets, we work out all possible mixing matrices and identify those which can
give reasonable leading order prediction. It turns out that 20 of the 27 cases do not contain any
element which can be identified as $S_{2\nu}$. Two of the remaining seven groups, namely
\group{162,10} and \group{486,125}, give predictions similar to the one that follows from \mt
symmetry, {\it i.e.} vanishing $\theta_{13}$ and maximal $\theta_{23}$. The other five groups
give reasonably good leading order predictions for the mixing angles. These are shown in Table II.
Noteworthy groups are \group{324,13} in category X and last two groups in the Table belonging to
category Y. These lead to the atmospheric angle in the 3$\sigma$ range of the best fit value and 
$\theta_{13}$ close to its 3$\sigma$ range. Thus it is conceivable that small perturbations in
these two case can lead the desired solution.

\section{Generating solar scale: Examples of perturbations}
\label{examples}
Our discussion so far was restricted to the leading order predictions for $\theta_{23}$ and
$\theta_{13}$ while $\theta_{12}$ was undefined due to the exact degeneracy in solar pair. We
now discuss possible perturbations which lift the degeneracy and hence generate the solar scale and
the required value for the solar angle. We find that a common symmetry
\be \label{s1tl}
S_1\equiv E=\left(\ba{ccc}
0&1&0\\
0&0&1\\
1&0&0\\
\ea \right) \ee
and different $T_l$ lead to the various predictions listed in Table \ref{SU3-results} in case of
$\Delta(6n^2)$ and $A_5$ groups. This is also applicable to the first two solutions obtained from
the group $PSL(2,7)$. Let us consider specific choices for $T_l$ for these
groups. The smallest $\Delta(6n^2)$ leading to $\sin^2\theta_{13}$ within
3$\sigma$ is $\Delta(384)$ in which case $S_1$ and 
\be
T_l=U(8,1,0)=-\left(
\ba{ccc}
0&1&0\\
\eta_8^*&0&0\\
0&0&\eta_8\\
\ea \right) \ee
with $\eta_8=e^{\frac{i\pi}{4}}$ leads to the prediction given in Table \ref{SU3-results}.
Analogous $T_l$ in case of $A_5$ and $PSL(2,7)$ are given by
\beqa \label{tla5}
T_l&\equiv& F(2,0,1)HE=\frac{1}{2}
\left(
\begin{array}{ccc}
 \mu _+ & -1 & \mu _- \\
 1 & -\mu _- & -\mu _+ \\
 -\mu _- & -\mu _+ & 1 \\
\end{array}
\right)~~{\rm for}~A_5~{\rm and}\nonumber \\
T_l&\equiv& NME=\frac{i}{\sqrt{7}} \left(
\begin{array}{ccc}
 \beta^4(\beta-\beta^6)   & \beta(\beta^4-\beta^3)& \beta^2(\beta^2-\beta^5) \\
 \beta^4(\beta^4-\beta^3) & \beta(\beta^2-\beta^5)& \beta^2(\beta-\beta^6) \\
 \beta^4(\beta^2-\beta^5) & \beta(\beta-\beta^6)  & \beta^2(\beta^4-\beta^3) \\
\end{array}
\right)~{\rm for}~PSL(2,L),\eeqa
where
$\mu_{\pm}=\frac{1}{2}(-1\pm\sqrt{5})$, $\beta=e^{2 \pi i/7}$ and $H$, $N$, $M$ are generators given
in \cite{Ludl:2010bj}.

The neutrino mass matrix invariant under $S_1$ in Eq. (\ref{s1tl}) has the structure
\be\label{mnu0}
M_{\nu}^0=m_0\left(\ba{ccc}
1&y&y\\
y&1&y\\
y&y&1\\
\ea \right).\ee
This form has been discussed as a leading order neutrino mass matrix in models based on the $S_3$
symmetry \cite{Jora:2006dh}. Here the same form arises due to the fact that $S_1$ is indeed an
element of the group $S_3$. Important difference is that $M_\nu^0$ above is not in the flavour
basis since $M_lM_l^\dagger$ invariant under $T_l$ is not diagonal. Thus leading order predictions
in two cases are entirely different.

\begin{figure}[!ht]
\centering
\subfigure{\includegraphics[width=0.4\textwidth]{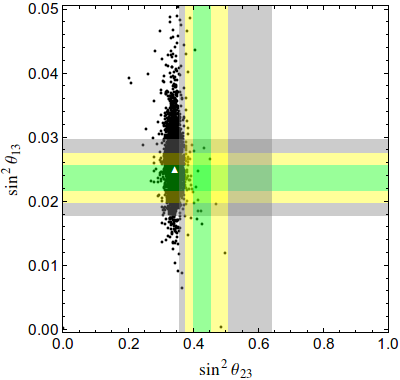}}\quad
\subfigure{\includegraphics[width=0.4\textwidth]{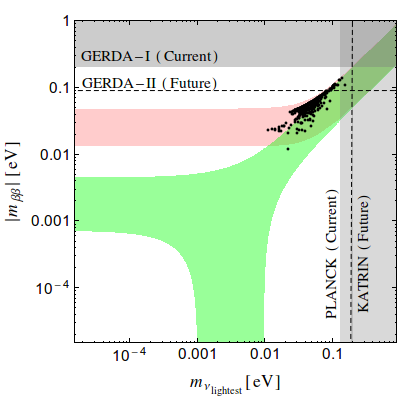}}
\caption{Results of the perturbations due to the breaking of $G_\nu$ in case of the group
$\Delta(6\cdot8^2)$. All the points in left panel are obtained for $|\epsilon_i|<0.05$ in Eq.
(\ref{mnup}) and correspond to the values of $\Delta m_{\rm solar}^2$, $\Delta m_{\rm
atm}^2$ and $\theta_{12}$ in agreement with the global fits at 3$\sigma$. Only points 
which fall in the $3\sigma$ bands in the left panel are shown in the right panel. The green,
yellow and gray bands in the left panel are the regions allowed at $1\sigma$, $2\sigma$ and
$3\sigma$ respectively by global fits. The white triangle is the prediction in the absence of
perturbations. The green (red) band in the right panel corresponds to the most general allowed
regions in case of normal (inverted) ordering in the neutrino masses. The gray
bands are the regions ruled out by the current strongest limits from GERDA-I
\cite{Agostini:2013mzu} and PLANCK and Galaxy clustering data \cite{Giusarma:2013pmn} while the
dashed lines correspond to the projected future limits by the ongoing experiments.}
\label{fig3}
\end{figure}

\begin{figure}[!ht]
\centering
\subfigure{\includegraphics[width=0.4\textwidth]{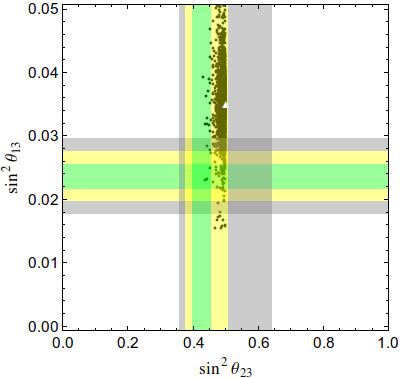}}\quad
\subfigure{\includegraphics[width=0.4\textwidth]{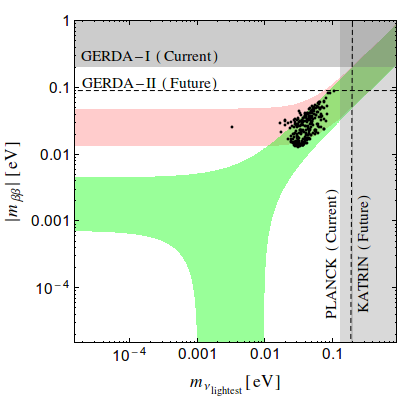}}
\caption{Same as Fig. \ref{fig3} but in the case of group $A_5$.}
\label{fig4}
\end{figure}

\begin{figure}[!ht]
\centering
\subfigure{\includegraphics[width=0.4\textwidth]{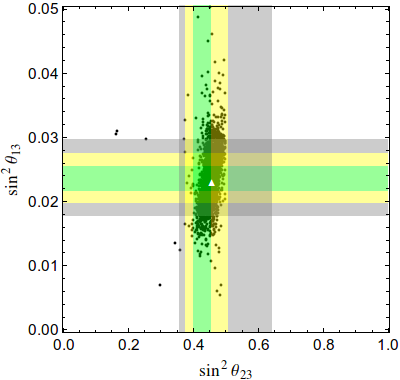}}\quad
\subfigure{\includegraphics[width=0.4\textwidth]{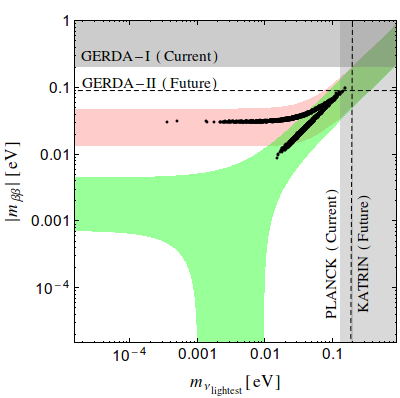}}
\caption{Same as Fig. \ref{fig3} but in the case of group $PSL(2,7)$.}
\label{fig5}
\end{figure}

The residual symmetry of neutrinos $G_\nu$ characterized by $S_1$ must be broken in order to
generate the splitting in solar neutrinos and corrections to $M_\nu^0$ may arise in various ways in
the realistic models. If such corrections are small, the modified neutrino mass matrix in the
presence of the  most general perturbations that may arise from the breaking of $G_\nu$ can suitably
be expressed
as
\be\label{mnup}
M_{\nu}=m_0\left(\ba{ccc}
1+\epsilon_1&y(1+\epsilon_3)&y(1+\epsilon_4)\\
y(1+\epsilon_3)&1+\epsilon_2&y\\
y(1+\epsilon_4)&y&1\\
\ea \right),\ee
where $\epsilon_i$ are in general complex parameters and $|\epsilon_i|\ll1$. In the diagonal basis
of the charged leptons, one obtains
\be \label{mnuf}
M_\nu^f = V_l^T M_\nu V_l, \ee
where $V_l$ is the unitary matrix which diagonalizes $M_lM_l^\dagger$ and is determined from
the underlying group using Eq. (\ref{talpha}) for specific $T_l$ as given above as long as  symmetry
corresponding to $T_l$ is unbroken. The corrected PMNS matrix can be obtained by diagonalizing the
above $M_\nu^f$ up to a freedom in interchanging the rows. The perturbations in $M_\nu$ modifies
the leading order predictions of mixing angles. We numerically study all three cases mentioned above
as they lead to reasonably good predictions at the leading order. For this, we randomly vary all the
parameters in $M_\nu^f$ considering them to be complex and evaluate the neutrino masses and mixing
angles for a
given $T_l$. The overall undetermined scale $m_0$ in Eq. (\ref{mnup}) is fixed by requiring the
correct atmospheric scale. The results are displayed in Fig. \ref{fig3} for the group
$\Delta(6\cdot8^2)$, in Fig. \ref{fig4} for the group $A_5$ and in Fig. \ref{fig5} for the group
$PSL(2,7)$. 
All the points in the left panels in the Figs. \ref{fig3}, \ref{fig4} and \ref{fig5} are obtained
for $|\epsilon_i|<0.05$ and requiring that they reproduce solar and atmospheric mass squared
differences and the solar mixing angle in the 3$\sigma$ ranges of their global fit values
\cite{Capozzi:2013csa}. Though the solar mixing angle is not predicted at the leading order by the
symmetries, even small $\epsilon_i$ can generate its required value because of the degeneracy in 
 solar pair. It can be seen from all the three figures that the reactor angle is more sensitive to
the small perturbations while the atmospheric mixing angle receives relatively small corrections.
Hence such corrections are more suitable for the group $A_5$ because leading order prediction of
$\theta_{13}$ gets sizable improvement as can be seen from Fig. \ref{fig4}. The corrections in
$\theta_{23}$ are however large enough to bring it in agreement with its global fit value at
3$\sigma$ in case of $\Delta(6\cdot8^2)$ group. In the right panels, we show the predictions for
effective mass of neutrinoless double beta decay $|m_{\beta \beta}|$ for the points in the left 
panels which reproduce both $\theta_{13}$ and $\theta_{23}$ in 3$\sigma$ ranges of their respective
global fit values. As can be seen, all three neutrinos remain quasidegenerate in the small
perturbation limit although original symmetry was designed to get only a degenerate solar
pair. Most points in the figure  fall in the region near to the sensitivity of the current
generation experiments. Large perturbations corresponding to $|\epsilon_i|>0.05$ further strengthen
the hierarchy in neutrino masses but lead to large deviations mainly in $\theta_{13}$. 

Similar results are shown in case of $PSL(2,7)$ in Fig. \ref{fig5} for which both the reactor
and atmospheric mixing angles are predicted within $1\sigma$ at the leading order. Small
perturbations
can generate the solar mixing angle and $\Delta m_{\rm solar}^2$ without introducing large
corrections in $\theta_{23}$ and $\theta_{13}$. Further, the neutrino mass spectrum can be more
hierarchical in this case compared to $\Delta(6\cdot8^2)$ and $A_5$.

\begin{figure}[!ht]
\centering
\subfigure{\includegraphics[width=0.4\textwidth]{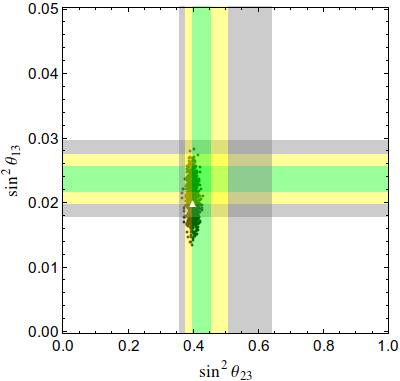}}\quad
\subfigure{\includegraphics[width=0.4\textwidth]{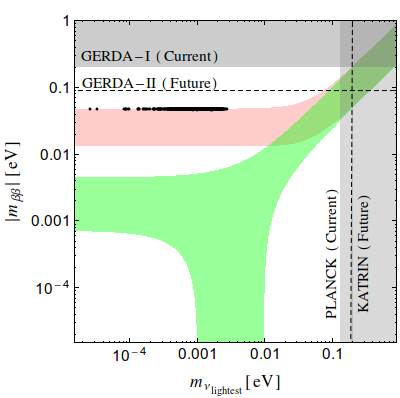}}
\caption{Same as Fig. \ref{fig3} but in the case of group \group{324,13}.}
\label{fig6}
\end{figure}

All three cases discussed above favor quasidegenerate neutrino spectrum and are consistent with both
normal and inverted ordering in the neutrino masses. Unlike this, the solutions based on the DSG of
$U(3)$ favor strict inverted hierarchy at the leading order. Small breaking of the residual
symmetry $G_\nu$ can then generate small mass for third neutrino and the splitting in the solar
pair. We discuss one such example based on the group \group{324,13} which offers the best
predictions at the leading order among all the DSG of $U(3)$. One of the possible choice of 
the residual symmetries of neutrinos and charged leptons in this case is
\be
\ba{cc}
S'_{1\nu}=\left(
\ba{ccc}
0&0&\eta_{12}\\
0&\eta_{12}&0\\
-i&0&0\\
\ea \right)~~{\rm and}~~
T'_l=-\left(
\ba{ccc}
0&\omega&0\\
0&0&1\\
1&0&0\\
\ea \right),
\\ \ea \ee
where $\eta_{12}=e^{2\pi i/12}$. The predicted leading order values of the mixing angles with this 
choice are given in Table \ref{U3-results}. The neutrino mass matrix with the above residual
symmetry and perturbed by the most general small corrections can be written as
\be\label{mnup1}
M'_{\nu}=m_0\left(\ba{ccc}
\epsilon_1&1&\epsilon_4\\
1&\epsilon_2&\eta_{12}^2(1+\epsilon_5)\\
\epsilon_4&\eta_{12}^2(1+\epsilon_5)&\epsilon_3\\
\ea \right).\ee
Numerical analysis is performed for $M'_\nu$ following the similar strategy described above and the
results are displayed in Fig. \ref{fig6}.
Unlike the examples based on the DSG of $SU(3)$, the preference for the strong inverted hierarchy is
clearly visible here. Apart from the solutions listed in the Table \ref{U3-results}, two the of the
$U(3)$ subgroups, namely \group{162,10} and \group{486,125}, lead to the \mt symmetric $M_\nu^f$ and
inverted hierarchy. The most general perturbations in this case are studied in \cite{Gupta:2013it}.
It is found that small perturbations can produce viable corrections in $\theta_{13}$ and
$\theta_{23}$ and both can be brought in to agreement with their global fit values in case of
quasidegenerate and inverted hierarchy in neutrino masses but not for the normal hierarchy.

\begin{figure}[!ht]
\centering
\subfigure{\includegraphics[width=0.4\textwidth]{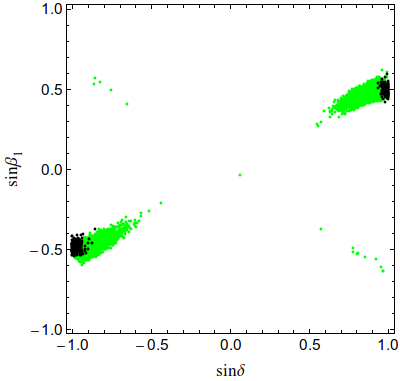}}\quad
\subfigure{\includegraphics[width=0.4\textwidth]{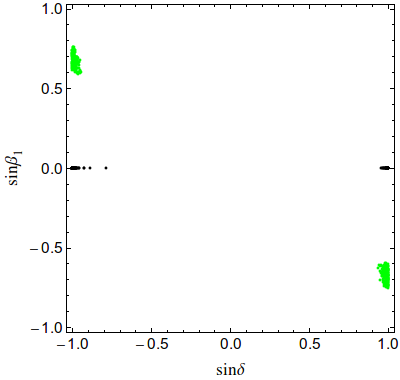}}
\caption{Correlations between the Dirac CP phase $\delta$ and Majorana phase $\beta_1$ associated
with quasidegenerate solar pair  arising from the small perturbations ({\it i.e.} $|\epsilon_i|\le
0.05$) in different cases. The black (green) points in left panel correspond to the  $\Delta(6\cdot
8^2)$ ($PSL(2,7)$) while the same in the right panel correspond to the case $A_5$ (\group{324,13}).
Only points that reproduce all three mixing angles in $3\sigma$ ranges of global fit are shown.}
\label{fig7}
\end{figure}

The CP violating phases $\delta$, $\beta_1$ do not remain arbitrary in all these cases once
perturbations are introduced. We have determined them numerically and results are shown in
Fig. \ref{fig7}. Only  points which reproduce all the mixing angles and solar and atmospheric mass
differences  in their 3$\sigma$ ranges are shown. One clearly sees a preference for the maximal CP
violation $\delta=\pm\frac{\pi}{2}$ in all the cases except  $PSL(2,7)$ which allows larger
deviation from it. We also estimate the invariant combinations of $\theta_{12}$, $\delta$ and
$\beta_1$ given in Eq. (\ref{correlations}). We find in all the cases except $PSL(2,7)$ that
perturbations only introduce small corrections to their values predicted at the leading order given
in Tables \ref{SU3-results}, \ref{U3-results}. As a result, the phases shown in Fig. \ref{fig7} are
consistent with the phases inferred from the analytic expressions given in Eq. (\ref{correlations})
and the leading order values of ${\cal I}_e$ and ${\cal I}_\mu$.  In the case of $PSL(2,7)$, the
perturbations lead to significant corrections in CP phases from their values predicted at the
leading order.

We discussed above the most general perturbation that may arise due to the breaking of $G_\nu$
and can be parametrized in terms of four complex parameters $\epsilon_i$ in Eq. (\ref{mnup})
and five in case of Eq. (\ref{mnup1}). Additional restrictions on $\epsilon_i$ can be imposed if
perturbation matrix $\delta M_\nu=M_\nu-M_\nu^0$ is assumed to posses some symmetry. For example,
following \cite{Joshipura:2014pqa}, it can be demanded that just like the leading order
contribution, the perturbation also arises from the spontaneous breaking of $G_f$. Thus, $\delta
M_\nu$ is invariant under a different symmetry $S_2$. Both $S_1$ and $S_2$ are assumed to be
subgroups of $G_f$ so that introduction of appropriate flavon fields can lead to both the pieces
after spontaneous breaking of $G_f$. Examples of such perturbations were given in
\cite{Joshipura:2014pqa} in the context of models discussed there. Let us give an economical example
here for the group \group{324,13}. Consider
the symmetry $S_2$
\be \label{s2}
S_2=\left(\ba{ccc}
-1&0&0\\
0&\eta_{12}^{10}&0\\
0&0&\eta_{12}^{10}\\
\ea \right)~.\ee
This $S_2$ is an element of the group \group{324,13}  and can be expressed as $S_2=S^2(12,3,7,3)$ in
terms of one of the generators of the group \group{324,13} as given by Ludl in \cite{Ludl:2010bj}.
Only the $(1,1)$ element of the  perturbation $\delta M_\nu$ invariant under $S_2$ is non-zero and
the resulting $M_\nu$ is thus obtained by putting $\epsilon_{2,3,4,5}=0$ in Eq. (\ref{mnup1}). The
simple three parameter neutrino mass matrix so obtained can  reproduce all the neutrino observables
within 1$\sigma$ with the choice of parameters 
$$m_0=0.0343~{\rm ~ eV}~~{\rm and}~~\epsilon_1= 0.0366 -0.0266~i.$$

\section{Summary}
\label{summary}
The recent progress in the field of discrete flavour symmetries suggest that such symmetries not
only can lead to the specific flavour mixing pattern in the leptonic sector but also can be used to
restrict the neutrino mass spectrum \cite{Joshipura:2013pga,Joshipura:2014pqa,Hernandez:2013vya}.
In this paper, we explored possible symmetries which can lead to the degenerate solar pair and a
massive or massless third neutrino together with a realistic leptonic mixing
pattern. Assuming that neutrinos are Majorana particles, it is shown that the DSG of $SU(3)$ can
lead to the former case while the latter can be achieved if the symmetry groups of the leptons are
the DSG of $U(3)$ and not $SU(3)$. The solar mixing angle becomes unphysical when the solar
neutrinos are exactly degenerate. It is show in this case that the symmetry groups determine the
atmospheric and reactor mixing angles and certain combinations of the elements of lepton mixing
matrix $U$, namely Im$(U_{\alpha1}^*U_{\alpha2})$ with $\alpha=e,\mu,\tau$. From pure group
theoretical considerations and utilizing the available information about the DSG of $U(3)$, we carry
out the detailed analysis for possible predictions of mixing angles and degenerate solar neutrinos
at the leading order. The study presented here encompass all the classified DSG of $SU(3)$ having
three dimensional IR. Our main observations are summarized in the following.
\begin{itemize}
\item None of the $SU(3)$ subgroups of type C leads to the realistic predictions for $\theta_{23}$
and $\theta_{13}$ simultaneously.
\item All the $SU(3)$ subgroups of type D predicts $\sin^2\theta_{23}\cos^2\theta_{13}=1/3$. In
this category, all the groups in $\Delta(6n^2)$ series for $n\ge8$ (except $10\le n \le 14$ and
$n=20,21$) can predict $\theta_{13}$ within the 3$\sigma$ range of global fit. The prediction for
$\theta_{13}$ by the other remaining groups in this category is subset of the predictions given by
the $\Delta(6n^2)$ groups.
\item Among the DSG of $SU(3)$ which are not in category C or D, the
group $PSL(2,7)\cong\Sigma(168)$ predicts both $\theta_{23}$ and $\theta_{13}$ very near to their
best fit values. 
\item Among all the DSG of $U(3)$ of order $<512$, only the group \group{324,13} predicts both
$\theta_{13}$ and $\theta_{23}$ simultaneously in their 3$\sigma$ ranges. This symmetry can lead to
a degenerate solar pair and a massless third neutrino as a leading order realization of inverted
hierarchy.
\end{itemize}

Suitable corrections to the above predictions are needed to generate the viable solar mass
difference and solar mixing angle. They may arise in complete models due to the breaking of residual
symmetries of neutrinos. The examples of such most general perturbations are discussed in the case
of $\Delta(6\cdot8^2)$, $A_5$, $PSL(2,7)$ and \group{324,13} groups and it is shown that small
perturbations can generate viable solar mass difference and all three mixing angles in agreement
with their global fit values. If perturbations are small, all three cases based on DSG of $SU(3)$
show preference for quasidegenerate neutrino spectrum while the group \group{324,13} predicts strong
inverted hierarchy. All these cases except $PSL(2,7)$ also lead to nearly maximal CP violating phase
$\delta$.

\begin{acknowledgments}
A.S.J. thanks the Department of Science and Technology, Government of India for support under the
J. C. Bose National Fellowship programme, grant no. SR/S2/JCB-31/2010. K.M.P. thanks the Department
of Physics and Astronomy of the University of Padova for its support. He also acknowledges partial
support from the European Union network FP7 ITN INVISIBLES (Marie Curie Actions,
PITN-GA-2011-289442).
\end{acknowledgments}

\bibliography{ref-pseudodirac.bib}
\end{document}